# On the Evolution, Numbers and Characteristics of Close-Binary Supersoft Sources

R. Di Stefano[1], L. A. Nelson[2]

[1] Harvard-Smithsonian Center for Astrophysics, Cambridge, MA 02138
[2] Bishop's University, Lennoxville, QC Canada, J1M 1Z7

**Abstract.** The ability to perform detailed evolutionary calculations is essential to the development of a well-defined and testable binary model. Unfortunately, traditional evolutionary calculations cannot be used to follow a significant fraction of possible close-binary supersoft sources (CBSSs). It is therefore important to examine the input physics carefully, to be sure that all relevant and potentially important physical processes are included. In this paper we continue a line of research begun last year, and explore the role that winds are expected to play in the evolution of CBSSs. We find that at least a subset of the systems that seemed to be candidates for common envelope evolution may survive, if radiation emitted by the white dwarf drives winds from the system. We study the effects of winds on the binary evolution of CBSSs, and compute the number and characteristics of CBSSs expected to be presently active in galaxies such as our own or M31.

## 1 Close-Binary Supersoft Sources

Close-binary supersoft sources (CBSSs) are binaries containing white dwarfs which can accrete matter from a more massive and possibly slightly evolved companion. Orbital periods during the epoch of supersoft source (SSS) behavior are typically on the order of one day. These systems are characterized by the fact that mass transfer occurs on a time scale closely related to the thermal time scale of the donor. The name "close-binary supersoft source" has been coined to distinguish these systems from other accreting white dwarf systems which may exhibit episodes of SSS behavior (see Tab. 1). The model was developed (van den Heuvel et al. 1992 [vdHBNR]; Rappaport, Di Stefano, & Smith 1994 [RDS]) as a possible description of systems such as CAL 83, CAL 87, and RX J0527.8–6954. (see Greiner et al. 1996, and Greiner, Hasinger & Kahabka (1991) for references.)

There are several features which make the CBSS model an attractive explanation for a subset of the observed luminous supersoft X-ray sources. Central among these is the fact that thermal-time-scale mass transfer from main-sequence or slightly evolved stars yields an accretion rate compatible with the steady burning of accreting hydrogen on the surface of a C-O white dwarf. The associated luminosity, temperature, and orbital period are in the range observed for $\sim 4-6$ of the sources (see Tab. 2). This is significant because no conventional white dwarf model (e.g., cataclysmic variables, or symbiotics) seems compatible with the systems' properties. Although neutron star (see Kylafis 1996 and

**Table 1.** Binaries that may appear as luminous supersoft X-ray sources

| (1) System Type | (2) Mass Transfer Mechanism | (3) $\dot{M}$ $M_\odot/\text{yr}$ | (4) $P_\text{orb}$ days | (5) Steady/ Recurrent | (6) Winds |
|---|---|---|---|---|---|
| CVs | mb &/or gr | $<\sim 10^{-8}$ | $<\sim 0.2$ | R* | nova** |
| CBSSs | thermal time scale readjustment of donor | $>\sim 10^{-7}$ | $\sim 0.2 - 3.0$ | S | yes |
| | | $<\sim 10^{-7}$ | $\sim 0.2 - \mathcal{O}(10^2)$ | R* | nova** |
| WBSSs | nuclear evolution of donor | $>\sim 10^{-7}$ | $\sim 3.0 - \mathcal{O}(10^2)$ | S | yes |
| | | $<\sim 10^{-7}$ | $\sim 3.0 - \mathcal{O}(10^2)$ | R* | yes |
| Symbiotics (Wind-Driven) | stellar winds from evolved donor | $>\sim 10^{-7}$ | $\mathcal{O}(10^2)$ | S | yes |
| | | $<\sim 10^{-7}$ | $\mathcal{O}(10^2)$ | R* | nova** |

(1) Here we consider only systems in which hydrogen-rich material is accreting onto the surface of a C-O white dwarf. CV= cataclysmic variable; CBSS=close-binary supersoft source; WBSS=wide-binary supersoft source; wind-driven symbiotics are the only binaries listed here in which the donor does not fill its Roche lobe. WBSSs are discussed further in Di Stefano (1996a,b). (2) The primary mass transfer mechanism is listed: mb=magnetic braking; gr= gravitational radiation. (3) $\dot{M}$ = rate at which mass impinges on the surface of the white dwarf; approximate values are given. (4) $P_\text{orb}$ = orbital period; approximate values are given. (5) Steady vs Recurrent SSS activity: when $\dot{M}$ is in the correct range (typically $>\sim 10^{-7} M_\odot/\text{yr}$) the source will burn nuclear fuel more-or-less steadily. * Recurrent sources: for smaller values of $\dot{M}$, hydrogen will burn sporadically. (6) Winds: "yes" indicates that the system is likely to emit a steady wind; ** "nova" indicates that mass ejection is likely to be primarily associated with nova explosions. Yungelson et al. (YLTTF; 1996) have performed a new population synthesis study that covers all of the binary systems listed in Tab. 1, save for wide-binary supersoft sources (WBSSs), which have been considered as a class more recently (Di Stefano 1996). In fact, YLTTF included in their analysis some systems not listed in Tab. 1, such as helium-accretors and planetary nebulae.

Hughes 1994) and black hole (see, e.g., Cowley et al. 1990) models have been suggested, there is no definitive evidence in favor of them for sources that emit only soft X-radiation. Thus, while keeping an open mind about the nature of the sources, and especially about the possibility that more than one physical model may be required to describe those systems whose nature has not yet been estab-lished, we choose to concentrate here on developing concrete, testable signatures of the white dwarf models.

## 2  The Importance of Evolution

This paper has two major themes. The first is the importance of developing an ability to carry out detailed evolutionary calculations. The second is the role played by winds in helping us to do this.

No binary model can be said to be well-defined or well-developed unless we know how to evolve the individual systems to which it is meant to apply. Yet, the standard approach to binary evolution proves to be problematic for many CBSSs. The reason for this is that the ratio $q$ of the mass of the donor, $m$, to that of the white dwarf accretor, $M$, is typically greater than unity. Thus, the donor's Roche lobe tends to shrink during mass transfer. If the donor cannot shrink at least as quickly, the evolution cannot be followed via the standard Roche-lobe-filling approach; in fact there is a real risk that a common envelope might form. Complicating matters is the fact that the donor in CBSSs is often so evolved that it is less able than a main sequence star of the same mass to shrink in response to mass loss. Thus, it appears that we may not be able to follow the evolution of a significant subset of CBSSs. It is therefore necessary to carefully consider whether all of the relevant physics has been included. As we will see in §3, we find that winds may play an important role in the physics and therefore also in the evolution.

The ability to track the evolution of specific systems transforms the conceptual CBSS model into a concrete model that is predictive and testable. It allows us to answer two types of question: (1) which of the observed systems may be realizations of the CBSS model? (2) what are the ranges of properties that a galactic population of CBSSs should be expected to exhibit?

### 2.1  Testing the Model for Individual Sources

Of the systems for which we have measured values of the orbital period, $P_{\rm orb}$, there are 6 whose properties are roughly consistent with the CBSS model. These are listed in Tab. 2. Of the systems listed, 3 (RX J0513.9–6951, CAL 83, and CAL 87) are strong candidates for the CBSS model. For each of these systems, the uncertainty boxes inferred for the bolometric luminosity and temperature enclose significant area within the steady-burning region computed by Iben (1982). Furthermore, the orbital periods are within the range computed by RDS for the CBSS model. But these circumstances are only weak arguments in favor of the model. To make a stronger argument, the model must make firm specific predictions for each system. Evolutionary calculations will allow us to compute from first principles $\dot{m}$, $\dot{M}$, and $\dot{m}_{ej}$, i.e., the donor's mass loss rate, the accretion rate of the white dwarf, and the rate of mass ejection from the system, respectively. We can then test whether these values are consistent with the observed luminosity and other properties of the system. Evolutionary calculations also allow us

**Table 2.** Observed Candidates for the Close-Binary Supersoft Model

| SSS | $P_{\rm orb}$ | $kT^*$ | $L$ (ergs/s) |
|---|---|---|---|
| RX J0513.9−6951 | 10.3 hrs | 30 − 40  (bb) | $0.1 - 6. \times 10^{38}$ |
| CAL 83 | 1.04 days | 20 − 50  (bb) | $0.6 - 2. \times 10^{37}$ |
| CAL 87 | 10.6 hrs | 65 − 75  (wd) | $1. - 10. \times 10^{38}$ |
| 1E 0035.4−7230 | 4.1 hrs | 40 − 50  (wd) | $0.8 - 2. \times 10^{37}$ |
| RX J0019.8+2156 | 15.8 hrs | 25 − 37  (wd) | $3. - 9. \times 10^{36}$ |
| RX J0925.7−4758 | 3.5 days | 45 − 55  (bb) | $1. - 10. \times 10^{37}$ |

These are the SSSs whose properties are not well-described by other accreting white dwarf models, but which may be roughly consistent with the CBSS model. *The fits used either a pure thermal model (bb) or a white dwarf model atmosphere model (wd). See Greiner 1996 for references.

to post-dict the prior history of each system, so that we can compute quantities such as the total mass ejected by the system throughout its evolution and the average luminosity. These post-dictions can also be checked for consistency with the data. For example, the study of the nebula surrounding CAL 83 seems to lead to a smaller estimate of the time-averaged luminosity over the past $\sim 10^5$ years than the present most-likely value of $L$ (Remillard, Rappaport & Macri 1994). We can determine if evolutionary calculations predict that the average value of $L$ over the past $\sim 10^5$ years is lower than the present value as determined by ROSAT. In general, evolutionary calculations allow us to better assess the likelihood that each observed system fits the CBSS model.

Three of the systems listed in Tab. 2 seem, on the face if it, either because of the value of their luminosity or of $P_{\rm orb}$, to be less likely members of the CBSS class. In two cases (particularly RX J0019.8+2156 and, to a lesser extent 1E 0035.4−7230), the inferred bolometric luminosities are low when compared to the position of the steady-burning region. But this does not mean that the systems are not CBSSs. It may mean that they are either on their way into or out of the steady-burning region. Computations of their possible evolutionary histories will help us to better constrain their nature.

### 2.2   Testing the Model for the Total Population: Past Work

RDS attempted to create, via a population synthesis analysis, the total population of close-binary supersoft sources that should be expected in the disk of our own Galaxy, or in other spirals, such as M31. They computed the properties of all sources whose luminosity and temperature fell squarely within the steady-burning region computed by Iben (1982). Their approach was conservative, and led to a prediction of $\sim 1000$ presently active CBSSs in the Galaxy. Later work by Di Stefano & Rappaport (1994) established that interstellar absorption would be expected to shield the vast majority of these systems from our view, so that the small number of sources observed in our Galaxy (∼ 6) and in M31 (∼ 15) is consistent with the large population computed by RDS.

Recently, a more comprehensive population synthesis study has been carried out by YLTTF. The new work is more complete in several ways. First, it includes all classes of SSS, except for wide-binary supersoft sources. Second, YLTTF compute the properties of CBSSs when the systems are not in the steady-burning regime, as well as when they are in it. YLTTF derive numbers that are similar to those computed by RDS. Given the uncertainties inherent in the calculations carried out by both groups, the ranges of numbers they compute are compatible. Since the two groups used somewhat different methods, the compatibility of their results seems to indicate that the results are robust.

Both RDS and YLTTF were hindered by the fact that complete evolutionary tracks could not be computed for a significant fraction of the systems created in their simulated galaxies. The two groups handled this difficulty in different ways, but each may be described as being appropriately conservative, in that they did not make specific predictions for the presence or observability of sources for which the standard evolutionary formalism was likely to fail.

## 3   A New Population Synthesis Study

Our goal is to perform a population synthesis to track the evolution of all systems which do not experience common envelopes during the phase in which mass is being transferred to the white dwarf. We differ from the previous population synthesis studies of CBSSs primarily in our approach to evolution. We include three new features. These are described sequentially in sections 3.1, 3.2, and 3.3. The evolutionary equations are presented in §3.4.

### 3.1   The Response of the Donor to Mass Loss

We specifically track, in each time step, the response of the donor to mass loss. In general, this response is described by the adiabatic index $\xi_{ad}$: $\xi_{ad} = d[log(r)]/d[log(m)]$, where $r$ is the radius of the donor, and $m$ is its mass. $\xi_{ad}$ can be estimated through comparisons with Henyey-like calculations. We find that a formula:

$$\xi_{ad} = \tilde{\xi}_{ad} \left[ 1 - \left( \frac{m_c}{\tilde{m}_c} \right)^2 \right], \qquad (1)$$

fits the numerical data, with the preferred values for $\tilde{\xi}_{ad}$ and $\tilde{m}_c$ equal to 4.0 and 0.2, respectively. Technically, $\tilde{m}_c$ depends on the value of $m$, but its dependence is relatively weak. Since the most massive cores found among donors in CBSSs are close to $0.2 M_\odot$, $\xi_{ad}$ can be small. Thus, CBSSs are at high risk for having the Roche-lobe-filling formalism break down.

In addition to this direct response, the donor may also be required to respond to the shrinking of its Roche lobe. In particular, if the donor can shrink to keep up with its Roche lobe, its radius may become smaller than the equilibrium radius of a star of the same mass and state of evolution. The star would therefore like to expand to achieve its equilibrium radius, $r_{eq}$, but can do so only on a thermal time scale. This thermal-time-scale push toward equilibrium introduces a further change in radius. We have

$$\frac{\dot{r}_{th}}{r} = \frac{r_{eq} - r}{f\,\tau_{KH}}, \qquad (2)$$

where $\tau_{KH}$ is the donor's Kelvin-Helmholz time, and the value of $f$ is determined by fitting to the results of Henyey-like calculations.

### 3.2 Mass Ejection By the White Dwarf

The white dwarf can retain only a fraction, $\beta$, of the total mass that falls onto its surface. That is, if the rate of accretion is $\dot{M}$, then only $\beta\dot{M}$ is burned and retained by the white dwarf. Because the accretor is a white dwarf, the study of steady nuclear burning provides some useful guidelines for the value of $\beta$ (see, for example, Paczyński 1970; Sion, Acierno & Tomcszyk 1979; Taam 1980; Nomoto 1982; Iben 1982; Fujimoto 1982; Fujimoto and Sugimoto 1982; Fujimoto & Taam 1982; Fujimoto and Truran 1982; Sion & Starrfield 1986; Livio, Prialnik & Regev 1989; Prialnik & Kovetz 1995). For each value of the mass ($M$) of the white dwarf, there is a range $\dot{M}_{\min} < \dot{M} < \dot{M}_{\max}$ within which matter that is accreted can burn more-or-less steadily as it accretes.

For $\dot{M} < \dot{M}_{\min}$, matter accumulates and burns explosively only after a critical amount of mass has been accreted. There are open questions about how much mass is ejected, but we find that these do not affect the numbers of sources dramatically. For $\dot{M}$ within the steady nuclear burning range, it is, in principle, possible that all of the incident mass can be burned. It was, however, pointed out that the start of nuclear burning is expected to turn on a wind, decreasing the maximum possible value of $\beta$ to something less than unity (Livio 1995). The case that has been least-well understood is that in which $\dot{M}$ is greater than $\dot{M}_{\max}$. The key question is whether accreting matter in excess of what can be burned accumulates in an envelope around the white dwarf, or whether it can be ejected "in real time". In the former case, the system risks undergoing a common envelope phase, unless the formation of a giant-like envelope can somehow turn off mass accretion when the white dwarf fills its own Roche lobe. Di Stefano et al. (1996b; DNLWR) made a case that, for some systems, the binary evolution could be stabilized if the excess matter can be ejected. The basic physics behind the argument is that (for $m > M$) the loss of mass from the system tends to allow the stars to move further apart and moderates the shrinkage of the Roche lobe, particularly if not much angular momentum is carried away by matter which leaves the system.

DNLWR showed that reasonable energy considerations alone would allow as much as $\sim 99\%$ of the incident matter to be ejected, if the rest of it was burned. Since then, Hachisu, Kato and Nomoto (1996) have verified that there are steady nuclear burning solutions in which excess matter is ejected in a steady wind. This justifies choosing the functional form of $\beta$ to fall as $\sim \dot{M}_{\rm max}/\dot{M}$ as $\dot{M}$ increases significantly above $\dot{M}_{\rm max}$ In the calculations used to derive the results presented in this paper, we have used the following prescription for $\beta$.

$$\dot{M} < \dot{M}_{\rm min}: \quad \beta = \Big(\frac{\dot{M}}{\dot{M}_{\rm min}}\Big)^{a_1}; \tag{3}$$

$$\dot{M}_{\rm min} < \dot{M} < \dot{M}_{\rm max}: \quad \beta = \beta_{\rm max} \tag{4}$$

$$\dot{M} > \dot{M}_{\rm max}: \quad \beta = \Big(\frac{\dot{M}_{\rm max}}{\dot{M}}\Big)^{a_2} \tag{5}$$

For our "standard" runs we chose $a_1 = 2$ and $a_2 = 1$. We also chose $\beta_{\rm max}$ to be unity; this is a conservative choice, almost certainly leading to an underestimate of the number of CBSSs which survive.

### 3.3 Radiation-Driven Winds

Because the white dwarf is the less massive of the companions, matter that leaves the system from its vicinity carries a large specific angular momentum. This is why radiation-driven winds, which can exit the system with smaller specific angular momentum, may play an important role. Typically the Roche-lobe geometry dictates that $\sim 5\%$ of the radiation emitted by the white dwarf strikes the donor. This generally represents $\mathcal{O}(100)$ times as much energy as is typically emitted by the donor. The radiation is soft, and does not penetrate deeply. It is therefore likely to generate a wind. Let $\dot{m}_{rd}$ represent the rate of mass loss due to a radiation-driven wind. In reality, only part of such a wind is likely to be emitted directly from the donor's surface, and part will be ejected from the highly irradiated accretion disk (see Popham & Di Stefano 1996). Thus, the angular momentum carried by the wind may be greater than that appropriate to the donor. Therefore, although in the work described here we will assume that the wind emanates from the donor, in ongoing work we relax this assumption.

### 3.4 The Evolutionary Equations

The Roche-lobe-filling condition, in combination with the principle of conservation of angular momentum leads to an expression for $\dot{m}$, the donor's total mass loss rate.

$$\dot{m}\mathcal{D} = \mathcal{N}. \tag{6}$$

In Eq. (6), $\mathcal{N} = m\big[\frac{\dot{J}_{dis}}{J} - \frac{1}{2}\frac{\dot{r}}{r}\big]$; $\dot{J}_{dis}$ is the rate at which angular momentum is drained from the system through the processes of gravitational radiation and magnetic braking, and $\dot{r}$ is the sum of the thermal term given by Eq. (2) and the donor's rate of change of radius due to nuclear burning. $\mathcal{D}$ is given by $\mathcal{A} + \beta \mathcal{B}$, and

$$\mathcal{A} = \left(1 + \frac{\xi_{ad}}{2}\right) - \frac{q}{2}\left\{\frac{1}{(1+q)} + \frac{f'}{f}\right\} - \frac{[\mathcal{F}q^2 + (1-\mathcal{F})]}{(1+q)} \qquad (7)$$

$$\mathcal{B} = \mathcal{F}\left\{-q + \frac{q}{2(1+q)} - \frac{q^2}{2}\frac{f'}{f} + \frac{q^2}{(1+q)}\right\} \qquad (8)$$

In these equations $\mathcal{F}$ is the fraction of the mass leaving the donor that is incident on the white dwarf; $\dot{M}$, the mass accretion rate of the white dwarf, is $\beta\mathcal{F}\dot{m}$. A generalization of these equations is considered in DiStefano 1996b.

## 4  Number of CBSSs in the Steady-Burning Region

To compute the numbers of CBSSs in the steady-burning region, we have performed a set of Monte-Carlo simulations. Each simulation was characterized by a different set of assumptions about the value of the common envelope ejection parameter, $\alpha$, and the distribution of properties among primordial binaries. Each simulation yielded a set of systems that might be expected to pass through a phase of mass transfer as a CBSS; we referred to these systems as CBSS candidates. We tracked the evolution of each CBSS candidate, and recorded its properties during the time it spent in the steady nuclear burning region. Because of uncertainties about how to perform the evolutions, each separate set of CBSS candidates was actually subjected to a variety of "treatments" in which different assumptions were made about the evolution (e.g., was there a radiation-driven wind? how efficiently could energy incident from the white dwarf be used to drive such a wind? etc.) Our preliminary results are shown in Tab. 3.

These results can be roughly understood from Fig. 1. The figure indicates that more systems can be evolved if there is a radiation-driven wind. Although this might lead us to think that there could be a significant increase in the number of systems in the steady-burning region, it is also the case that radiation-driven winds tend to moderate the mass accretion rate. Thus, some systems that might have been steady nuclear burners will not now enter the steady nuclear burning region; others may spend less time there. The overall number of SSSs that should be active in a galaxy such as our own at any one time will therefore not be radically different in calculations which include the effects of winds.

## 5  Properties of CBSSs in the Steady-Burning Region

Although the numbers of systems are not very different from what has been found in previous investigations, there are differences in the distribution of system properties. The most profound difference is that we find that winds will be a prominent feature of many CBSSs. Winds were not included in either the early (RDS) or more recent (YLTTF) population synthesis studies of these systems. Indeed since the donors have evolved to, at most, the base of the giant branch

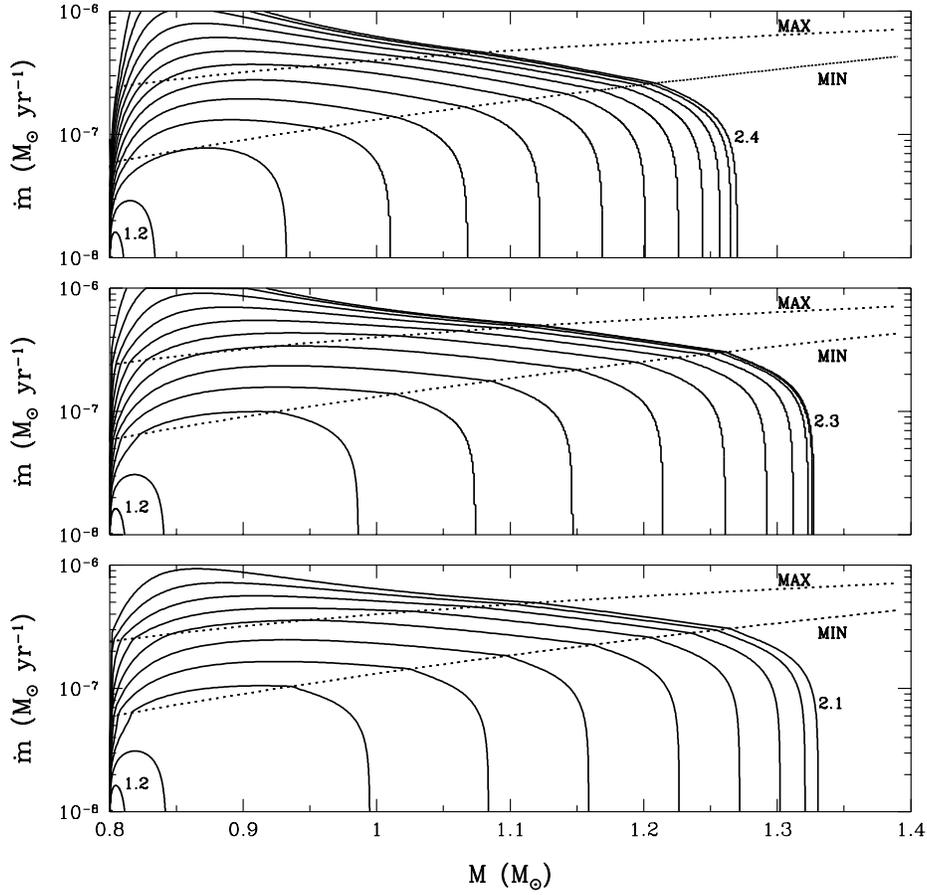

**Fig. 1.** The mass of the white dwarf is plotted along the horizontal axis; the rate at which mass is falling onto the surface of the white dwarf is plotted along the vertical axis. In all of the evolutionary tracks shown, the initial white-dwarf mass is taken to be $0.8 M_\odot$, and the donor starts on the TAMS. Within each panel, each curve is characterized by the initial value of the donor's mass, $m_d(0)$; the bottom curve has $m_d(0)$ marked; $m_d(0)$ is incremented by 0.1 in each subsequent curve. The value of $m_d(0)$ associated with the upper curve of each panel is the largest value in the sequence for which $\mathcal{D} > 0$ throughout the evolution. No radiation-driven winds were included in the evolutions shown in the bottom panel; radiation-driven winds, with an efficiency factor of 0.001 (0.01) were included in the middle (top) panel. Note that low to moderate efficiency can be expected to maximize both the number of active sources and the number of possible supernovae. As the wind efficiency increases, only slightly more systems can be evolved, but the mass transfer rates tend to be lower than would be necessary for significant mass accretion to occur.

**Table 3.** Numbers of CBSSs in the Steady-Burning Region

| $Treatment^*$ | 1 | 2 | 3 | 4 | 5 |
|---|---|---|---|---|---|
| Data Set 1** | | | | | |
| $f = 0$ | $\sim 2400$ | $\sim 3200$ | $\sim 2600$ | $\sim 2900$ | $\sim 2500$ |
| $f = 0.001$ | $\sim 1400$ | $\sim 1700$ | $\sim 1200$ | $\sim 4000$ | $-$ |
| Data Set 2 | | | | | |
| $f = 0.0001$ | $-$ | $\sim 1200$ | $-$ | $-$ | $-$ |
| $f = 0.001$ | $-$ | $\sim 1500$ | $-$ | $-$ | $-$ |
| $f = 0.01$ | $\sim 500$ | $\sim 1500$ | $\sim 4400$ | $\sim 2500$ | $-$ |

* "Treatment" refers to the values chosen for $\tilde{\xi}_{ad}$, $a_1$, and $a_2$. Treatment 2 is our "standard" treatment: $\tilde{\xi}_{ad} = 4.0$, $a_1 = 2$, $a_2 = 1$. Treatment 1 is more conservative in that $\tilde{\xi}_{ad} = 2.0$. Treatment 3 tends to increase the number of system that can be evolved, since the specific angular momentum of all mass leaving the system is chosen to be that appropriate to the donor; however, it also moderates the mass transfer rate, so that some systems that would otherwise be steady nuclear burners do not enter the steady nuclear burning region. In treatment 4, $a_1 = 1$; this allows for more mass to be accreted below the steady nuclear burning region. Treatment 5 is an "optimal treatment; the same parameters are used as for the standard treatment, but $\mathcal{D}$ is set to a minimum value whenever the computed value dips below zero. ** Data set 2 is our "standard" data set; the population synthesis calculation that gave rise to it had $\alpha$, the common envelope efficiency factor, set equal to 0.8. To derive data set 1, $\alpha$ was taken to be 0.1. The wind efficiency factor, $f$, is a constant of proportionality that enters into the relationship between $\dot{M}$ and the rate at which matter is lost through the radiation-driven wind. Details can be found in DN; here, the values can be viewed as providing a relative measure of the efficiency of producing the wind; there is no radiation-driven wind for $f = 0$.

during the phase of most-active mass transfer, they would not be expected to emit a large wind if they existed in isolation. However, the white dwarf's ejection of matter that it cannot burn in "real time", coupled with radiation-driven winds, will lead many CBSSs to eject a significant wind. Thus, systems in which there has been significant mass ejection may, in essence, shield themselves from our view. YLTTF have considered this process for wind-driven symbiotics. Investigations of the distribution of properties among the systems we expect to be able to detect must therefore explicitly consider winds ejected from the system, and their ability to shield active systems from our view. Such investigations are underway.

## 6  Conclusions

It has long been known that the evolution of a binary in which the donor is more massive than the accretor, and also possibly slightly evolved, can be problematic. ROSAT's discovery of the class of luminous supersoft X-ray sources, and the subsequent proposal of the close-binary supersoft (CBSS) model, has forced us to face the difficulties associated with such potentially problematic evolutions. Indeed the motivation for doing so is strong, since detailed evolutionary calculations are required to shape the CBSS model into a well-defined theory which is testable both for individual observed sources and for the range of characteristics one should expect among a galactic population of sources.

If nature has thus presented us with a problem to solve, it has also been kind in choosing to pose the problem in a venue that contains some obvious clues to its resolution. The fact that the accretor is a white dwarf is important, because the physics of nuclear burning places useful constraints on the value of the mass retention factor, $\beta$, as a function of the white dwarf mass and accretion rate. Further, if, as recent work suggests, $\beta$ can become small when the accretion rate exceeds the rate compatible with steady nuclear burning, this helps to stabilize the evolution. The white dwarf nature of the accretor also leads to radiation-driven winds. Indeed, an important feature of the work described here, which will be presented in more detail elsewhere (Di Stefano 1996b; Di Stefano & Nelson 1996), is the inclusion of radiation-driven winds. It is physically reasonable to expect such winds because of (1) the tremendous energy associated with nuclear burning, and (2) the fact that the radiation emitted by the white dwarf is so soft that it cannot penetrate beyond the outer layers of the donor. Mass ejection through radiation-driven winds can help to stabilize the binary evolution. Furthermore, winds are a feature in some of the observed systems which may be close-binary supersoft sources (see Greiner 1996 for references).

We find that, across a broad range of assumptions about the evolution of the binary systems both prior to and during any CBSS phase, the number of steady-nuclear burning sources is in the range of $\sim 1000-4000$ presently active systems. The inclusion of radiation-driven winds does not have a dramatic effect on the number of presently active close-binary supersoft sources. Some systems that would have entered the steady-burning region if winds had not been included, will not do so now; others, which could not be evolved before, will live in the steady-burning region during part of their active mass-transfer phase. What is perhaps more interesting are some subtle shifts in the distribution of system properties; we are presently investigating these. An important observational consequence of the work presented here is that it predicts that winds are likely to be ejected from many, perhaps most, CBSSs during their transit across the steady-burning region. Further, the total amount of mass ejected from the system from the earliest stages of mass transfer up until the present can be calculated.

Results that go beyond those presented here, both in the extent of the parameter space explored and in the analysis of the distribution of system properties, will be described in DN.

*Acknowledgement:* We would like to thank Scott Kenyon, Saul Rappaport, and J. Craig Wheeler for discussions, and Trevor Wood for assistance with calculations performed during the early phases of the work. This work has been supported in part by NSF under GER-9450087 and by the NSERC (Canada).